 \documentstyle[epsfig,preprint,aps]{revtex} 
 
\begin{document}
\draft

 
\title
{The Net Spin Model}
 
\author{ H. Q. Lin$^{1}$, J. L. Shen$^{1,2}$, and H. Y. Shik$^{1}$ }
\address{
$^{1}$Department of Physics, Chinese University of Hong Kong, Hong Kong, China\\
$^{2}$State Key Laboratory of Magnetism,
     Institute of Physics, Chinese Academy of Science, Beijing, China
}
 
\date{\today}

\maketitle

\begin{abstract}
We study a set of exactly soluble spin models in one and two dimensions
for any spin $S$. Its ground state, the excitation spectrum, quantum phase
transition points, as well as dimensional crossover are determined.
\end{abstract}


\newcommand{\beq}{\begin{equation}}
\newcommand{\eeq}{\end{equation}}
\newcommand{\beqn}{\begin{eqnarray}}
\newcommand{\eeqn}{\end{eqnarray}}
\newcommand{\beqx}{\begin{eqnarray*}}
\newcommand{\eeqx}{\end{eqnarray*}}
\newcommand{\bsigma}{\mbox{\boldmath $\sigma$}}

It is well known that a spin-1/2 antiferromagnetic chain with nearest-neighbor
coupling $J_1$ and next neighbor coupling $J_2$ could dimerize in the presence
of frustration, as exemplified by the exactly soluble one-dimensional
Majumdar-Ghosh (MG) model \cite{mg_model},
where the exact (twofold degenerate) ground
state is a simple product of singlet dimers. The elementary excitation can be
constructed as a pair of unbound spins above the completely dimerized state
\cite{shastry-sutherland}. This model also can be considered as a two-chain
with one diagonal or a zigzag spin ladder. 
Analytical and numerical studies of the MG model (or models alike)
\cite{mg_study}
show a transition from a gapless
phase when $J_2 < J_{2c}$ to a gapped phase when $J_2 > J_{2c}$.
In addition, a family of spin-ladder models were studied by Kolezhuk and
Mikeska \cite{kolezhuk-mikeska} that exhibit non-Haldane spin-liquid
properties as predicted by Nersesyan and Tsvelik \cite{nersesyan-tsvelik}.
Great progress were made in the studies of
the ladder systems \cite{Dagotto-Rice}.
In two dimensions, Shastry and Sutherland proposed a model (SS)
with dimerized eigenstate twenty years ago \cite{Shastry-Sutherland2}.
The model had its experimental realization in compound
SrCu$_2$(BO$_3$)$_2$ recently \cite{srcubo_1}.
Theoretical and experimental investigations of the SS model are currently
undergoing \cite{srcubo_2}.

In this Letter we consider a set of quantum spin models,
we call it the net spin model,
with any spin $S$ defined on a double layer as shown in Fig. 1.
Each layer has $M \times L$ sites and it connects to the other
not only perpendicularly by coupling $J_1$,
but also diagonally by $J_3$ and $J_4$.
$L$ is a measure of the dimensionality 
varying from 1 (1D, the net spin ladder) to $M$ (2D, the net spin layer).
Throughout this work, we use $\bf S$ to represent spins on the lower layer,
and $\bf S'$ on the upper layer.
Open boundary conditions are imposed.

Under certain combination of coupling constants,
we show that for general $L$ and $M$ and any spin $S$
the model could be solved exactly for a subset of its Hilbert space.
First, depending on parameters, the model have two kinds of ground state,
one is the completely dimerized state,
and the other one is the ground state of the spin-$\sigma$ ($\sigma=2S$)
model defined on the single layer.
Second, the model exhibits rich excitation phases, depending on coupling
constants and dimensionality.
One may think of the net spin model as a {\it generalization of
the Majumdar-Ghosh model for any spin $S$ in two dimensions},
with much more rich phase diagrams than that of the MG model.

The net spin model Hamiltonian is:
\begin{eqnarray}
H &=& 2J_1 \sum_{k,l=1}^{M,L}   {\bf S}_{k,l}  \cdot {\bf S'}_{k,l}
\nonumber \\
  &+& 2J_2 \hspace{-0.3cm} \sum_{k,l=1}^{M-1,L-1} \hspace{-0.3cm}
        {\bf S}_{k,l}  \cdot ( {\bf S}_{k+1,l}  + {\bf S}_{k,l+1} )
      + {\bf S'}_{k,l} \cdot ( {\bf S'}_{k+1,l} + {\bf S'}_{k,l+1} )
        \nonumber \\
  &+& 2J_3 \sum_{k,l=1}^{M-1,L-1}
        {\bf S'}_{k,l} \cdot ( {\bf S}_{k+1,l}  + {\bf S}_{k,l+1} ) \nonumber\\
  &+& 2J_4 \sum_{k,l=1}^{M-1,L-1}
        {\bf S }_{k,l} \cdot ( {\bf S'}_{k+1,l} + {\bf S'}_{k,l+1} ) ~,
\label{model_h}
\end{eqnarray}
with $J_1, J_2, J_3, J_4 \geq 0$, and $|{\bf S}| = |{\bf S'}| = S$.
Both $J_1$ and $J_2$ favor local antiferromagnetic ordering 
while $J_3$ and $J_4$ represent frustration effects.

For any integers $M$ and $L$, we found that a complete dimerized state:
\beq
   \psi_D = [1,1'][2,2'] \cdots [M,M'] \cdots [N,N'] ,~  (N=ML)
\label{dimer_conf}
\eeq
is an eigenstate of Eq. (\ref{model_h}) when
\beq
   2 J_2 = J_3 + J_4 ~.
\label{dimer_cond}
\eeq
To prove our statement,
we first show that this is true for any 4-spin plaquette,
$\{ {\bf S}_{1,1}, {\bf S'}_{1,1}, {\bf S}_{2,1}, {\bf S'}_{2,1} \}$,
simplified as $\{ {\bf S}_1, {\bf S'}_{1}, {\bf S}_2, {\bf S'}_{2} \}$,
as indicated in Fig. 1.
Since $\psi_D$ is a direct product of singlet $[1,1']$ and $[2,2']$, we have
\beq ( {\bf S}_{1} + {\bf S'}_{1} ) \psi_D =
     ( {\bf S}_{2} + {\bf S'}_{2} ) \psi_D = 0 ~, \eeq
\beqn H &=& J_1  \left[~ 
   ( {\bf S}_{1} + {\bf S'}_{1} )^2 + ( {\bf S}_{2} + {\bf S'}_{2} )^2
 - ( {\bf S}_{1}^2 + {\bf S'}_{1}^2 + {\bf S}_{2}^2 + {\bf S'}_{2}^2 )
		~\right] \nonumber \\
&+&2J_2 ( {\bf S}_{1} + {\bf S'}_{1} ) \cdot ( {\bf S}_{2} + {\bf S'}_{2} )
 + (J_3 + J_4 - 2J_2) ( {\bf S}_{2} \cdot {\bf S'}_{1}
		      + {\bf S}_{1} \cdot {\bf S'}_{2} ) ~. \eeqn
Immediately we see that dimerized state $\psi_D$ is an eigenstate of $H$ with
eigenvalue $-4J_1 S(S+1)$ when condition Eq. (\ref{dimer_cond}) holds.

In Eq. (\ref{dimer_conf}), $[i,j]$ denotes the normalized wave function of
spin singlet $|{\bf S}_i + {\bf S}_j| = 0$, 
with Clebsch-Gordon coefficients for spin-$S$ dimer given by:
\beq <m, -m | 0, 0> = { (-1)^{S-m} \over \sqrt{2S+1} } ,
m = S, S-1, \cdots, -S ~. \eeq

We know that the formation of dimers $[1,1']$ and $[2,2']$ is due to
antiferromagnetic coupling $J_1$.
Without antiferromagnetic couplings $J_3$ and $J_4$, quantum
fluctuation due to $J_2$ will kill the formation of both dimers.
As long as condition Eq. (\ref{dimer_cond}) holds, the effect of $J_2$ on
the dimers will be cancelled out exactly by $J_3$ and $J_4$.
Thus, no matter how many plaquettes we put together to form
the two dimensional lattice as shown in Fig. 1, $\psi_D$ is the eigenstate
of Hamiltonian Eq. (\ref{model_h}) with eigenvalue $E_D = - 2 J_1 N S(S+1), N=LM$.

It is also obvious that the dimerized state $\psi_D$ is the ground
state in the limit of $J_1 \rightarrow \infty$, and
in the limit of $J_1 \rightarrow 0$, $\psi_D$ is not the ground state,
so there exists a critical value $J_{1c}$
such that when $J_1 > J_{1c}$ the ground state is completely dimerized.

We should mention that for the case of $L=1, S=1/2$,
the net spin model (the net spin ladder) goes back to
the MG model \cite{mg_model}
when $J_2 = 0.5 J_1, J_3 = J_1$, and $J_4 = 0$.
{\it So our model is a generalization of the MG model for any spin $S$.}
Moreover, our model applies to two dimensional geometry and its ground state
and the first excited state are exactly soluble for a wide range of parameters.

To study the quantum phase transition and the excitation spectrum of the net
spin model, we concentrate on two cases, 
one is $L=1$, corresponding to the one-dimensional lattice,
and the other one is $L=M$, corresponding to the two-dimensional lattice.
We present qualitative discussions for cases $1 < L < M$ at the end.
For the purpose of illustration, we take $J_3=J_4=J_2$ so that it
enables us to get analytical solutions.
Such simplification does not alter general features of the model,
although to get exact numbers, one needs to perform numerical calculations
for other combination of parameters $\{ {J_2,J_3,J_4} \}$.
For $L=1$,
several studies on the generalization of the MG model,
\cite{shastry-sutherland,mg_study,net_ladder},
were carried out recently.

We start from the ladder case, $L=1$, the Hamiltonian can be rewritten as
\beq
H = - 2M J_1 S(S+1) + J_1 \sum_{k=1}^{M} {\bsigma}_{k}^2
  +  2J_2 \sum_{k=1}^{M-1} {\bsigma}_{k} \cdot {\bsigma}_{k+1} ~,
\label{model_sone}
\eeq
where ${\bsigma}_{k} = {\bf S}_{k}   + {\bf S'}_{k}$.
This Hamiltonian describes a chain of $M$ spins with 
$|{\bsigma}_{k}|$ range from 0 (singlet, dimer), 1 (triplet), $\cdots$, to
the highest angular momentum $2S$.

With this Hamiltonian,
it is easy to see that $\psi_D$ is an eigenstate of $H$ by using the relation
$( {\bf S}_{k} + {\bf S'}_{k} )[k, k'] = 0$, hence
\beq
 \left( J_1 \sum_{k=1}^{M} {\bsigma}_{k}^2
      + 2J_2 \sum_{k=1}^{M-1} {\bsigma}_{k} \cdot {\bsigma}_{k+1}
 \right) |\psi_D> = 0 ~,
\eeq
and the eigenvalue is $E_D = - 2M J_1 S(S+1)$.

To lower the energy, the squared term, $J_1 {\bsigma}_{k}^2$,
favors the system to be in the $|{\bsigma}_{k}| = 0$ dimer state,
while the exchange term, $J_2 {\bsigma}_{k} \cdot {\bsigma}_{k+1}$,
favors the system to be in the higher $|{\bsigma}_{k}|$ state.
If all pairs ${\bf S}_{k} + {\bf S'}_{k}$ are in the same angular momentum
$\sigma$ state, then the model
is equivalent to the spin-$\sigma$ chain with coupling constant $2J_2$,
apart from a constant proportional to $MJ_1$.
One can see from Eq. (\ref{model_sone}) that the eigenstates of
the spin-$\sigma$ chain is also the eigenstate of the net spin ladder model.
In fact, as $J_1$ decreases further, the ground state of the net spin ladder
model becomes the ground state of the spin-$\sigma$ chain.
The transition point $(J_2/J_1)_c$ is determined by
\beq
   \langle \sum_{k=1}^{M} {\bsigma}_{k}^2
   + 2 \left( {J_2 \over J_1} \right)_c 
   \sum_{k=1}^{M-1} {\bsigma}_{k} \cdot {\bsigma}_{k+1} \rangle = 0 ~,
\label{j2overj1c}
\eeq
where the expectation value is taken in the ground state of
the spin-$\sigma$ chain.
To evaluate the ground state energy of the spin-$\sigma$ model, 
we observed that the linear spin wave theory \cite{Anderson-LSW} gives
the ground state energy of the spin chain within 3\% accuracy as compared
with exact calculations such as the exact diagonalization \cite{hql} and
the density matrix renormalization group method \cite{White-Huse}.
So even though it is inappropriate to describe the magnetic properties
such as the staggered magnetization in one and two dimensions,
it is perfectly safe to use the spin wave theory for
the estimation of the transition points $(J_2/J_1)_c$, given by
\beq
   \left( {J_2 \over J_1} \right)_c
	= {\pi \over 2} {\sigma + 1 \over \pi ( \sigma + 1) - 2} ~.
\label{Jc1d}
\eeq
For spin $S = 1/2$ and 1,
we have verified those results by exact diagonalization calculations.

Furthermore, we can show that once $(J_2/J_1) > (J_2/J_1)_c$,
the ground state jumps from all $\sigma_k=0$ (the dimer state)
to all $\sigma_k=2S$, without passing any intermediate values of $\sigma_k$.
The transition is of the first order.

We next study the excitation spectrum of the model.
%
Let us establish an important fact first.
Instead of using $\{ {\bf S}_{k}, {\bf S'}_{k}, k=1, 2, \cdots, N=LM \}$
as basis operators, one can also use
$\{ {\bsigma}_{k} = {\bf S}_{k} + {\bf S'}_{k}, k=1, 2, \cdots, N=LM \}$
to specify eigenstates.
The complete dimerization state would be written as
$\psi_D = [0] [0] \cdots [0] ~.$
Then, one can show that 
$  \psi_m = [0] \cdots [0] [m] [0] \cdots [0] ~,$
and
$  \psi_{m,m'} = [0] \cdots [0] [m] [0] [m'] \cdots [0] ~,$
and wave functions consisting nonzero $\sigma$s separated by at least one dimer
($\sigma_k=0$), are also eigenstates of the Hamiltonian Eq. (\ref{model_sone}).

For sufficient large value of $J_1$, the ground state is the complete
dimerized state and one can create a single magnon simply
by changing one $\sigma_k$ from $0$ to $1$.
For example, for spin-1/2 case, 
\beq
   \psi_m = [1,1'][2,2'] \cdots (\uparrow, \uparrow)\cdots [M,M'] ~,
\label{psim}
\eeq
is the eigenstate of the net spin model with total spin $S_{tot}=1$.
The energy gap in this case is exactly $\Delta_{st} = 2J_1$ and
the transition is from singlet to triplet.
This excitation energy is independent of other antiferromagnetic couplings
as long as they satisfy the condition Eq. (\ref{dimer_cond}) and
thus it is possible that there exist states with lower excitation energy.

Indeed, as we have shown in detail for the spin-1/2 case \cite{hql-shen},
when $(J_2/J_1) > 1/2$,
there exists another singlet with energy lower than the triplet.
The corresponding wave function,
defined as $\psi_{\Box}$ with eigenvalue $E_s = J_1 - 4J_2$,
is that all pairs are dimerized except
on one plaquette where the two dimers are broken into two triplets
to make up a non-dimer singlet.
This state is also highly degenerate for this plaquette could be anywhere.
We can show that when $2J_2 > J_1$, two isolated triplets embedded in
otherwise all dimers would have higher energy than
that of combining the two triplets next to each other to form a singlet.
In another words, there exists an attractive interaction 
between the two triplets.
Consequently, as more and more triplet excitations are created, 
a kind of phase separation would occur where the system consists of
clusters of triplet and singlet.
Note that this critical value $(J_2/J_1)=1/2$ is independent of lattice size
and dimensionality.

For the general spin-$S$ case, same conclusions hold. 
Using the fact that $\psi_{m}$ ($E_t$), $\psi_{\Box}$($E_s$), 
$\psi_{m,m'}$ ($2E_t$), etc.,
are also eigenvectors, we can show that
\beqn
   \Delta_{st} &=& E_t - E_D = 2J_1 \\
   \Delta_{ss} &=& E_s - E_D = 4(J_1 - J_2) \\
   \Delta_{ss} &<& \Delta_{st}, \hspace{1.0cm} J_2/J_1 > 1/2 ~.
\eeqn
To be more specific, we use $S=1$ as an example and summarize our findings
in Figure 2a. There are three regions, depending on parameter $J_2/J_1$:

\begin{description}
\item[Region I], $J_2/J_1 \leq 1/2$, $\Delta/J_1= \Delta_{st}/J_1 = 2$,
independent of $J_2$ and it is singlet to triplet.

\item[Region II], $1/2 < J_2/J_1 < (J_2/J_1)_c$,
$\Delta/J_1 = \Delta_{ss}/J_1 = 4 ( 1 - J_2/J_1 )$ decreases linearly from
$2$ to $4 ( 1 - (J_2/J_1){c} ) = 1.4760$ at $(J_2/J_1)_c = 0.6310$.
The excitation is from singlet (dimers) to another singlet.
In both regions I and II, the completely dimerized state $\psi_D$ is
the ground state.

\item[Region III], $J_2/J_1 > (J_2/J_1)_c$,
the ground state is no longer the completely dimerized state $\psi_D$.
Rather, it is the ground state of spin $\sigma=2$ chain with coupling $2J_2$,
degenerate with $\psi_D$ right at $(J_2/J_1)_{c}$.
The excitation is the Haldane gap of the spin-2 chain,
$\Delta/J_1 \approx 0.088$ \cite{spin-2}
If the original spin is $S$, then in this region the ground state is that
of the spin-$2S$ chain. For large $S$, the Haldane gap is approximately
$\Delta \approx 9.5 S^2 e^{- \pi S}$.
\end{description}


We now turn to the double layer case ($L=M$ and $N=LM$).
Similarly, we get
\begin{eqnarray}
H &=& - 2N J_1 S(S+1) + J_1 \sum_{k=1}^N {\bsigma}_{k}^2
   + 2J_2 \sum_{<k,l>}^N {\bsigma}_{k} \cdot {\bsigma}_{l} ~,
\label{model_s12d}
\end{eqnarray}
where ${\bsigma}_{k} = {\bf S}_{k,l}   + {\bf S'}_{k,l}$
and $<k,l>$ refers to nearest neighbors in two dimensions.
Again, when all $|{\bsigma}_k|$ are the same,
this is nothing but the two-dimensional spin-$\sigma$ antiferromagnetic
Heisenberg model with coupling constant $2J_2$, apart from a constant.
The critical point at which the complete dimerized state becomes the
excited state is determined by,
\beq
\sigma ( \sigma + 1 ) = - \left ( {J_2 \over J_1} \right )_c {2 \over N}
   \langle \sum_{<k,l>}^N {\bsigma}_{k} \cdot {\bsigma}_{l} \rangle ~.
\eeq
and the spin wave theory gives
\beq
   \left( {J_2 \over J_1} \right )_c
	= {1 \over 2} {\sigma + 1 \over 2 \sigma + 0.3159} ~.
\label{Jc2d}
\eeq
As one increases $J_2$, similar to the chain case,
the change of $\sigma_k$ in the ground state is directly 
from $\sigma_k=0$ to $\sigma_k=2S$, no intermediate values of $\sigma_k$ appear. 


For the excitation spectrum, again we can show that it is still true
when $J_2/J_1 \leq 1/2$, the first excitation is a triplet. 
However, for the layer case, without detailed calculation of the ground
state energy of the corresponding spin-$\sigma$ AFH model,
one can easily show that $(J_2/J_1)_c < 1/2$.
This is because on the 2D square lattice, due to quantum fluctuations,
the true ground state energy is always lower than that of
the N\'{e}el state, $-2N \sigma^2$.
Moreover, when $(J_2/J_1) > (J_2/J_1)_c$, the system is gapless because
the existence of long-range-order in the 2D square lattice spin-$\sigma$
Heisenberg model for $\sigma \geq 1$,
as shown rigorously long time ago \cite{dyson-et,neves-perez}.
Therefore, there exists no transition from the dimer singlet
to another singlet as what we have seen for the net spin ladder.
We plot the excitation gap as function of $J_2/J_1$ in Figure 2b.

What will happen for cases of $1 < L < \infty$?
Exact estimation of the ground state energy of spin-$\sigma$ model is not
necessary. 
As we have shown, whether the excitation spectrum will be 1D or 2D like 
depends on whether $(J_2/J_1)_c >$ or $< 1/2$.
According to the linear spin wave theory, we can write
\beq \langle \sum_{<k,l>} {\bsigma}_{k} \cdot {\bsigma}_{l} \rangle
   = -a \sigma^2 - b\sigma ~, 
\left ( {J_2 \over J_1} \right )_c
   = {1 \over 2} {\sigma + 1 \over a \sigma + b} ~.
\eeq
where $a=2-1/L$ accounts for the N\'{e}el state energy,
and $b$ ranges from 0.3634 (1D) to 0.3159 (2D).
Exact account of the quantum fluctuation will not change this number much.
We can safely use these numbers to make estimations.
For spin $S=1/2$, we have studied $L=2$ and $L=3$ cases \cite{hql-shen}
by using the spin wave theory, perturbation theory, and
exact diagonalization on small lattices, they are all consistent.
Results indicate that there seems to have a dimensional crossover 
from 1D to 2D occurred at $L=3$ to $L=4$.

For other combinations of couplings, one can only determine
the critical points and eigenvalues of the model numerically
(see, for example Ref. \onlinecite{num_ladder}).
However, qualitative behaviors should be the same as what we have shown here.
Note that the energy gap $2J_1$ is quite large (critical values of $J_2$
are smaller than $J_1$) so the phase diagram should be similar to Fig. 2.
Furthermore, for the double layer case with spin $S=1/2$, we have shown
that in the net spin model,
the critical value is $J_3 = J_4 = J_2 = 0.4288 J_1$.
While other studies of the ``standard'' double layer antiferromagnetic
Heisenberg model (corresponding to $J_3 = J_4 = 0$ in the net spin model),
obtained the critical value as
$J_2/J_1 = 0.3960 \pm 0.0003$ \cite{Shevchenko-et}.
These two values are quite close, suggesting that
our results are insensitive to the values of $J_3$ and $J_4$.
Essential physics is determined by the two couplings: $J_2$ and $J_1$.

In summary,
we have studied a class of quantum spin models, the net spin model,
defined by the Hamiltonian Eq. (\ref{model_h}) for any spin $S$.
The model is controlled by three antiferromagnetic couplings:
$J_2/J_1$, $J_3/J_1$, $J_4/J_1$, and the measure of the dimensionality $L$.
When $2 J_2 =  J_3 + J_4 $,
the completely dimerized state $\psi_D$ is an eigenstate of the model
and it is also the ground state if $J_1$ is sufficiently large.
Moreover, wave functions such as $\psi_{m,m'}$ and $\psi_{\Box}$
are also eigenstates of the model.
So a subset of the Hilbert space of the net spin model is exactly known.
To fully understand the model,
we have analytically studied the case of $J_2 = J_3 = J_4$ extensively
and obtained phase diagrams for both 1D (ladder, $L=1$)
and 2D(double layer, $L=M$) cases.
For the ladder case, when
$S=1/2, L=1, J_2 = 0.5 J_1$, $J_3 = J_1$, and $J_4 = 0$
the model goes back to the Majumdar-Ghosh model \cite{mg_model}.
Thus, the net spin model is a generalization of the MG model for
any spin $S$ in two dimensions.
We also showed qualitative different behavior between 1D and 2D,
and discussed the crossover from 1D to 2D.

We acknowledge
I. Affleck,
C.D. Batista,
D.K. Campbell,
A.H.C. Neto,
G. Ortiz,
and S.A. Trugman
for stimulating discussions.
This work was supported in part by the Earmarked Grant for Research from
the Research Grants Council (RGC) of the HKSAR.


\end{document}